\documentclass[12pt]{article}
\newlength{\dinwidth}
\newlength{\dinmargin}
\usepackage{epsf}
\usepackage[intlimits,tbtags]{amsmath}
\usepackage{amsfonts}
\usepackage[section]{placeins}
\usepackage{subfigure}
\usepackage[small,bf]{caption2}

\begin{document}
\setcounter{page}{0}
\newcommand{\be}{\begin{equation}}
\newcommand{\ee}{\end{equation}}
\newcommand{\br}{\begin{eqnarray}}
\newcommand{\er}{\end{eqnarray}}
\newcommand{\lp}{\left(}
\newcommand{\rp}{\right)}
\newcommand{\lk}{\left\{}
\newcommand{\rk}{\right\}}
\newcommand{\lc}{\left[}
\newcommand{\rc}{\right]}
\newcommand{\sT}{{\scriptscriptstyle T}}
\newcommand{\2}{\,2}
\newcommand{\dif}{\mathrm{d}}
\def\a{\alpha}
\def\b{\beta}
\def\g{\gamma}
\def\l{\lambda}
\newcommand{\se}{\section}
\newcommand{\tra}{\vec{p}_{\sT}}
\newcommand{\Z}{Z\left(\beta\right)}
\newcommand{\half}{\frac{1}{2}}
\newcommand{\ninteger}{\left.\mathbb{Z}\right.^n}
\newcommand{\nreal}{\left.\mathbb{R}\right.^n}
\newcommand{\omegat}{\widetilde{\Omega}}
\numberwithin{equation}{section}
\thispagestyle{empty}

\begin{flushright}
FFUOV-04/17
\end{flushright}

\vspace*{2cm}

{\vbox{\centerline{{\Large{\bf Thermodynamic nonextensivity in a 
closed string gas}}}}}

\vskip30pt

\centerline{Manuel A. Cobas, M.A.R. Osorio, Mar\'{\i}a Su\'arez
\footnote{E-mail addresses: cobas, osorio, maria@string1.ciencias.uniovi.es}}

\vskip6pt
\centerline{{\it Dpto. de F\'{\i}sica, Universidad de Oviedo}}
\centerline{{\it Avda. Calvo Sotelo 18}}
\centerline{{\it E-33007 Oviedo, Asturias, Spain}}

\vskip .5in

\begin{center}
{\bf Abstract}
\end{center}


Well known results in string thermodynamics show that there is always a
negative specific heat phase in the microcanonical description of a gas of
closed free strings whenever there are no winding modes present. We will
carefully compute the number of strings in the gas to show how this negative
specific heat is related to the fact that the system does not have
thermodynamic extensivity. We will also discuss the consequences for a
system of having a microcanonical negative specific heat versus the exact
result that such a thing cannot happen in any canonical (fixed temperature)
description.

\newpage

\section{Introduction and foundations}

Unclear progress has been made during the last  years in order to
unambiguously settle the longstanding problem posed by Hagedorn \cite{hage}
 about the statistical behavior of strings. It is established that, in
general, there is not equivalence between the canonical (fixed temperature)
and the microcanonical (fixed energy) descriptions. This non equivalence is
related, although not equivalent, to the appearance of a negative specific
heat in the microcanonical description. This fact has been used by some
authors as a sort of selection principle to discard any model with a
microcanonical negative specific heat regime. We will comment on this issue
in the conclusions to mainly remind to the reader that it is only in the
canonical descriptions that the specific heat cannot be negative.
Microcanonical negative specific heats are always related to a thermodynamic
non extensivity of the system. More precisely, in terms of the entropy, the
entropy of the system is not the raw sum of the entropies of the independent
subsystems when equilibrium among them is supposed to exist. This special
behavior may be interpreted as a sign of a first order phase transition, but
can also be taken as an indication of the appearance of a non equilibrium
state. Anyhow, what seems to be the key ingredient for closed strings is the
fact that the negative specific region covers the whole high energy regime.
This happens, at least, as long as no meaning can be found to
$\beta$-duality and its corresponding version in terms of energy.

On the other hand, this thermodynamic non extensivity for the Hagedorn gas
is related to the well known image in which a single fat string absorbs
energy thorough its vibrational modes preventing the increasing of the
temperature that gets closer to the Hagedorn one as long as energy
increases.  What we give here, among other things, is an explicit
calculation of the number of strings in the particular case in which there
is no winding mode state in the gas. We use the bosonic string living in 25
big dimensions as the tool instead of, for example, the SSTII, simply
because its high dimensionality makes easier the convergence of some plots
we present in our work.

To begin with, to properly make Thermodynamics, one should put the
system inside a sort of a box as it is done with a classical system of
particles. We do this for strings in a fundamental way by taking a particular
compact geometry for the space. A limit like the high volume limit has to be
understood and physically defined over this finite volume picture as it is
done for the analogous point-particle system. Now, we know more clearly that
this is a subtle point when extended objects as strings are involved
\cite{nuestro,otros}.  Other important aspect to remark is that  the
treatments usually called canonical (see for instance \cite{fraut}), are
really macrocanonical. The number of strings in the gas is a function of the
temperature and the volume and not a free variable.  Indeed, it is in this
ensemble where the implementation of a null chemical potential is more
natural, because
$\mu$ is a variable of the grand partition function. When classical
Maxwell-Boltzmann (MB) statistical counting is applicable, it is easy to
implement this equilibrium condition over the canonical ensemble too. Things
are far more difficult when quantum statistics is involved because, as it is
well known, it is hard to get an explicit expression for the canonical
partition function and it is really troublesome, it seems that analytically
impossible, to compute, in a closed form, the derivative of the Helmholtz free
energy with respect to $N$, the number of constituents.

In Statistical Mechanics, when $\Gamma\lp E,V,1\rp$,
 the number of one object accessible states of a system, is much bigger than
the number $N$ of objects, the subtleties of Bose-Einstein (BE) or
Fermi-Dirac (FD) statistics become superfluous; it is highly improbable for
two objects to try to get into the same state since there are lots of rooms
to accommodate into. From $\Gamma\lp E,V,1\rp$, by taking one derivative
with respect to the energy, one can obtain the single-object density of
states $\Omega\lp E,V,1\rp$ and from it it is easy to obtain the single
object partition function as $q=Z\lp\beta,V,1\rp=
\sum_{E}\,\Omega\lp E,V,1\rp\,{\mathrm e}^{-\b\,E}$ where $\b=1/T$ being $T$
the temperature in units in which the Boltzmann constant $k$ equals one. When
use is made of MB counting, the canonical partition function for $N$
independent objects can be written as $Z\lp \beta,V,N\rp=q^N/N!$. This is
really an approximate calculation valid for free (independent) objects or, in
a more realistic manner, dilute gases at sufficiently high temperatures. The
macrocanonical description rests upon the grand partition function $\Theta\lp
\b,V,\l\rp=\sum_{N}\, Z\lp \beta,V,N\rp\,\l^N= \sum_{N}\,\l^N\, \lc
q\lp\beta,V\rp\rc^N/N!$, where $\l$ is finally identified with
$\mathrm{e}^{\,\b\mu}$ and the second equality holds as long as MB statistics is
a good approximation. When physical conditions are such that $\mu=0\,\,(\l=1)$
we get $\Theta\lp \b,V,1\rp =\sum_{N}\,\lc q\lp\beta, V\rp\rc^N/N!=
\mathrm{e}^{\,q}$, again when MB statistics is applicable. Imposing
$-\beta\mu=\partial\,\mathrm{ln}\,Z\lp
\beta,V,N\rp/\partial N =0$ gives, for big $N$ that is what one needs to
make thermodynamics, $\overline{N}=Z\lp \beta,V,1\rp$ as the number of
objects for the system at a given temperature $T$ when MB statistics is
applicable. When BE or FD statistics are necessarily involved, the grand
canonical partition function is the best starting point to face the more
complicated combinatorics, although actual computations can finally be
written in terms of $q$.

Our main goal here is to face the problem of the black-body radiation
looking at the system as one for which energy is conserved. The property of
the system to have null chemical potential, when one assumes that, at least
at low energies, quantum statistics is relevant, makes useful the use of a
system for which the chemical potential is fixed (to zero in our case) and
the number of objects depends finally on energy and volume. The ensemble
adapted to these conditions has a partition function given by
$\Lambda\lp E,V,\mu\rp = \sum_{N}\,\Omega\lp E, V, N\rp\,\lambda^N$ that,
for $\mu=0$, gives $\Lambda\lp E,V,0\rp =\sum_{N}\,\Omega\lp E, V, N\rp$. Here
$\Omega\lp E, V, N\rp$ is the microcanonical density of states for the
system with $N$ objects. Thermodynamics can be gotten from $\beta H$ where
$H$ is the enthalpy $E+PV$ with $P$ the pressure. Namely, $H=E+PV=TS+N\mu$.
Since in our case $\mu=0$, we have that $\beta H=S=\mathrm{ln}\,\Lambda\lp
E,V,0\rp$. The computation of the average number of objects then gives

\be
\overline{N}\lp E,V,\lambda=1\rp = \lc \lambda \lp \partial
\,\mathrm{ln}\,\Lambda/\partial\,\lambda\rp_{V,E}\rc_{\lambda=1} 
= \frac{\sum_{N=0}^{\infty}\,N\,\Omega\lp
E,V,N\rp}{\sum_{N=0}^{\infty}\,\Omega\lp E,V,N\rp}
\label{nbar1}
\ee
If it is assumed that, in the summations appearing in the numerator and the
denominator, there is a value of the label $N$ equal to $N^*$ such that each
sum can be well approximated by the  single $N^*$ contribution, we get that
$\overline{N}=N^*$. Assuming that $N^*$ is the same for the sum in the
numerator and the denominator is equivalent to assuming that no fluctuations
in the number of objects are present. Quantitatively,
$N^*$ is such that $\lc\partial\,\Omega/\partial\,N\rc_{N=N^*}=0$ which is
just  the number of particles we get by imposing $\mu=0$ in the (genuine)
microcanonical ensemble characterized by the density of states $\Omega\lp
E,V,N\rp$. In other words, we assume the equivalence of the microcanonical
and the "enthalpic" thermodynamical descriptions based on the fact that
$S=\mathrm{ln}\,\,\Lambda\lp
E,V,0\rp=\mathrm{ln}\,\sum_{N=0}^{\infty}\,\Omega\lp E,V,N\rp\approx
\mathrm{ln}\,\Omega\lp E,V,N^*\rp$.

It is also interesting to try to express the average number of
objects in terms of the single object  partition function $q\equiv
Z_1\lp\beta\rp$ (it is a function of temperature and volume although only its
dependence thorough $\beta$ will be kept explicit for convenience as  it
will happen for other functions. The result is

\be
\overline{N}(E)= \frac{\sum_{n=0}^{\infty}\,\,{\cal
L}^{-1}\,\left\{\sum_{r=1}^{\infty}\,\lp \pm\rp^{r+1}\,Z_1\lp r\beta\rp
\right\}\,\ast \widetilde{\Omega}_n\lp E\rp}{\sum_{n=0}^{\infty}\,
\widetilde{\Omega}_n\lp E\rp}
\label{nbar}
\ee
where $\mathcal{L}^{-1}$ means inverse Laplace transformation
with respect to the variable $\beta$  to $E$, $\widetilde{\Omega}_n\lp
E\rp\equiv (1/n!)\mathcal{L}^{-1}\left\{\lc\sum_{r=1}^{\infty}\,\lp\pm\rp^{r+1}\,Z_1\lp
r\beta\rp/r\rc^n\right\}$, the star  means convolution and, in the sums over
$r$, the plus sign corresponds to BE statistics and the minus sign to FD
statistics\footnote{Here: $\widetilde{\Omega}_n\lp E\rp
=(1/n!)\,\mathcal{L}^{-1}\lk \mathrm{ln}^n\, \Theta\lp \b, V,
1\rp\rk$, with $\mathrm{ln}\, \Theta\lp \b, V, \lambda\rp = \pm \sum_i
\,\mathrm{ln}\,\lp 1\pm\lambda\,\mathrm{e}^{-\b \epsilon_i}\rp 
= \sum_{r=1}^{\infty}\,(\mp)^{r+1}\lambda^{r}\frac{Z_1\lp \beta r\rp}{r}$. FD 
statistics corresponds to the upper sign.}. Everything comes from rewriting the entropy as

\be
\Lambda\lp E, V, 0\rp = \mathcal{L}^{-1}\lk\Theta\lp \beta, V, 1\rp\rk =
\sum_{n=0}^{\infty}\,\widetilde{\Omega}_n\lp E\rp
\ee
where $\widetilde{\Omega}_0\lp E\rp= \delta\lp E\rp$ which corresponds to
the fact that there is only one state with zero objects and it is the one
with zero energy. It is very important to remark that, for example,
$\widetilde{\Omega}_1\lp E\rp$ is not the single object density of states
$\Omega\lp E,V,1\rp = \mathcal{L}^{-1}\lk Z_1\lp\beta\rp\rk$. 
For instance, using BE statistics, what $\widetilde{\Omega}_1\lp E\rp$
really does is to take into account all the possibilities for allocating
several bosons in a way in which all of them are in the same state. The
counting is then made in terms of clusters or bunches of bosons (or in terms
of the number of different filled quantum states). This can be seen if we
look at the way of computing the canonical partition function with quantum
statistics. One can write

\be
Z\lp \beta, N\rp = \sum_{\lk n_i, \lambda_i\rk}\,\prod_i\,\frac{\lc
Z_1(n_i\beta)\rc^{\lambda_i}}{n_i^{\lambda_i}\,\lambda_i!}
\label{canonical}
\ee 
where the sum is restricted over the set of positive integer values
$\lambda_i$, $n_i$ such that $\sum_i\,\lambda_i\,n_i = N$. From
\eqref{canonical} we get that the contribution to the canonical partition
function $Z\lp\beta,N\rp$ coming from having the $N$ objects in the same
quantum state is given by $Z_1\lp N\beta\rp/N$. Now, if we sum over the
number of objects $N$ from one to infinity we get
$\sum_{N=1}^{\infty}\,Z_1\lp N\beta\rp/N$ whose inverse Laplace transform
just gives $\widetilde{\Omega}_1\lp E\rp$ for BE counting. As an example, 
$\widetilde{\Omega}_2\lp E\rp$ gives the density of states for a system of
2-clusters, i.e., two objects in different sates, three objects distributed
two of them in the same state and a single object in a different one, and,
in general, $N$ objects allocated in a manner such that only two different
single object states are occupied. $\Omega \lp E, V, N\rp$ and
$\widetilde{\Omega}_{n=N}\lp E\rp$ only coincide when classical (MB) statistics
is applied because then, as explained at the beginning of this introduction,
$Z_{MB}\lp \b, N\rp =q^N/N!$ and so $\widetilde{\Omega}_n=\mathcal{L}^{-1}\lk
q^n/n!\rk=\Omega_{MB}\lp E,n\rp$. If one is interested in knowing the
dependence on the volume, one finds that, when $q$ depends linearly on the
volume, $\Omega\lp E, V, N\rp$ goes as $A_1\,V^N + A_2\,V^{N-1}+...+
A_N\,V$ as long as $\widetilde{\Omega}_N \sim V^N$. Anyhow, everything
finally fits in place because, in our case, what is relevant to physics is
the sum over $N$. One can then reorder the sum regrouping the terms
proportional to $V^N$ and so get the same sum over the label $n$ of
$\widetilde{\Omega}_n$ instead of the sum over the number of particles $N$
of $\Omega_N$.

One of the goals of this work is to show that the number of strings, as a
function of energy and volume, can be computed even when quantum statistics
has to be taken into account (we will see that this would be the case for a
low number of open spatial dimensions).  In particular, we will exemplify
thorough the gas of closed bosonic strings that the number of strings tends
toward a constant value $N_H$ as long as energy increases once the
microcanonical negative specific heat phase has been reached. To try to make a
clear exposition, section two will be devoted to present a microcanonical
description of the traditional blackbody radiation problem. In section
three, we will present the problem of computing thermodynamic quantities for
a gas of strings that has a big number of open spatial dimensions; this will
be exemplified thorough a gas of bosonic strings, although the results
clearly hold for other closed strings under the same conditions. After the
computation of $\overline{N}(E,V)$, emphasis will be made on the non
extensivity of the entropy $S(E, V)$ at high energy . Finally, we will
present a discussion about what can be said when there is a negative
specific heat in a microcanonical approach, the equivalence of
microcanonical and canonical descriptions and under which conditions a
negative specific heat can be used as a kind of selection rule to say that a
system is unphysical.

\section{A microcanonical description of the blackbody
radiation problem}

The blackbody radiation problem is a Statistical Mechanics classic. 
Although the problem is posed in such a way that a microcanonical (fixed
energy) description should be used, this is not the standard procedure in
text books. Here we adopt such point of view stressing the fact that an
"enthalpic" description in terms of an ensemble in which the chemical
potential $\mu$ is fixed and null (at a given energy and volume) would be
the most natural first attempt to describe that  system.  Eq. \eqref{nbar}
gives now

\be
\overline{N}(E; d) = \frac{\zeta\lp d\rp}{\zeta\lp
d+1\rp}\,\frac{\sum_{n=1}^{\infty}\,n\,\widetilde{\Omega}_n^{BE}\lp
E\rp}{\sum_{n=0}^{\infty}\,\widetilde{\Omega}_n^{BE}\lp E\rp}
\label{ental}
\ee
$\widetilde{\Omega}_n^{BE}\lp E\rp$ is easily
calculated to give

\be
\widetilde{\Omega}_n^{BE}\lp E; d\rp =
\frac{a_0^n}{\Gamma\lp dn\rp\,n!}\,\lc\frac{V\,\zeta\lp
d+1\rp\,\Gamma\lp d\rp}{2^{d-1}\,\pi^{d/2}\,\Gamma\lp d/2\rp}\rc^n\,E^{\,dn-1}
\ee
both functions are expressed, for later convenience, as depending on the
number of space dimensions, $d$. The factor of $a_0^n$ comes from the
contribution of the spin degrees of freedom for a massless object and it will
equal $2^n$ for a  photon gas. The next step to really make \eqref{ental}
useful is to approximate the sums in the numerator and the denominator by the
contribution of an $n^{*}$  such that the logarithm of each sum is dominated
by the logarithm of that term $\Omega_{n^*}$. The terms of maximum
contribution for the numerator and the denominator only differ in ${\mathrm
ln}\,n$ which is negligible as compared to $n$ when $n$ is big. If one uses
that $n^*$ is big and Stirling's approximation, one gets that

\be
n^*_{BE}\lp E; d\rp = \lc \frac{a_0\,d^{-d}\,\zeta\lp d+1\rp\Gamma\lp
d\rp}{2^{\,d-1}\,\pi^{d/2}\,\Gamma\lp d/2\rp}\,V\,E^d\rc^{1/\lp d+1 \rp}
\ee

For the photon gas with $a_0=2$ in three space dimensions one exactly gets

\be 
\overline{N}(E;3) = \frac{\zeta\lp 3\rp}{\zeta\lp 4\rp}\,n^*_{BE}\lp
E;3\rp 
\ee
that, of course, coincides with the value given by the macrocanonical
description once the expression of $n^*_{BE}$ is substituted.

The ensemble average inverse temperature can be computed for the photon gas
along the same lines to give

\be
\overline{\b} = \lc\frac{\partial S}{\partial E}\rc_{V} =
\frac{3\,n^*_{BE}\lp E;3\rp}{E}=3\,\frac{\zeta\lp 4\rp}{\zeta\lp
3\rp}\frac{\overline{N}}{E}
\ee
It is interesting to emphasize that $n^*_{BE}$ is not the average number of
particles because $\zeta\lp 3\rp\neq\zeta\lp 4\rp$. In fact, this difference
is what gives a quantum correction to the classical ideal gas equation of
state. It may also be instructive for treating fermionic strings to get the
same results in the enthalpic description for a gas of massless fermions at
$\mu = 0$. For them, the main basic ingredient is the single-cluster density
given by

\be
\widetilde{\Omega}_1^{FD}\lp E\rp = \lp 1 -
\frac{1}{2^{\,d}}\rp\,\widetilde{\Omega}_1^{BE}
\ee
As for bosons, from here, after repeated convolutions one may easily get
$\widetilde{\Omega}_n^{FD}\lp E\rp$. 
From the results for bosons and fermions it is easy to get the
thermodynamics for the supersymmetric system.

Before ending this section, some comments come into place. The connection
between Statistical Mechanics and Thermodynamics rests upon the fact that the
number which connects atomic Physics and the scale of grams is Avogadro's
number; a big number indeed. Consequently, many
approximations as the use of Stirling's formula over $\mathrm{ln}\,N!$ are
based on the big number of objects involved. Here, however, what has been
assumed to be a big number is $n^*$ which refers to the number of clusters or
different occupied states. Assuming that the main contributions come from big
$n$'s, we are then supposing that one gets the most relevant contributions to the
entropy from the situations in which a big number of objects have different
quantum numbers. The fermionic and supersymmetric calculations show that
there will be no difference between bosons and fermions when the space
dimension grows big. With a big number of dimensions there would be a lot of
room to move in and, consequently, a big number of accessible states even
when energy is not too high.

\section{The stringy blackbody}

The approach to the problem of a system of strings in equilibrium such that
the chemical potential is null has been widely treated in numerous
references in the past. In many cases, the question about what ensemble is
used to describe the problem has not been clear. This way, descriptions that
should be called grand canonical have been named canonical at the same time
it was sustained that, at least from the beginning, a quantum description in
terms of bosons and fermions were in course. Perhaps this is, in some cases,
a remnant of the presentation made by classical text books in Statistical
Mechanics when treating the gas of photons (cf. \cite{KHuang}). Even though
$\mu$ cannot be computed in the quantum canonical description, we know that
the equivalence of this picture and the one gotten from an ensemble in which
the chemical potential is fixed and not an averaged quantity will rests upon
the possibility of substituting the sum over the number of objects in the
grand canonical description by the contribution of a preponderant single
term $N=N^{\ast}$. This is equivalent to assuming that there are no big
fluctuations in the number of objects and this, if true, may happen for a
picture in which temperature is fixed or for one in which energy is given
and not an averaged quantity. It has been exposed in the introduction that,
it is only after taking something like the dilute gas approximation to make
proper use of MB statistics, when one can certainly compute in a canonical
description the number of objects that corresponds to a vanishing chemical
potential and compare it to the more natural description in terms of an
ensemble with fixed and vanishing chemical potential and an averaged number
of objects. In fact, when making the approximation of a free theory, what
finally plays the most important role for the MB approximation to hold is
the number of big dimensions one finally has.

By assuming that this is the physical picture to make thermodynamics, a
fundamental question arises about whether a decompactified description in
which no winding mode contribution appears can be reached from a particular
finite volume description for the gas by taking an infinite volume limit.
That question can be resolved in affirmative if one realizes that the mass
of a winding mode is proportional to the radius. Then, there must exist a
radius big enough so as to get that, for a given energy, the masses of the
oscillators (proportional to a positive integer number) are much lighter
than the mass of the winding modes. The degeneration of the oscillators
increases much faster than that of the winding modes and this is the
explanation for the negative specific heat in the oscillator ruled high
energy phase and this is what a decompactified phase for the string gas
really means. Among other things, we are going to study a standard bosonic
string gas to see the characteristics of the transition from the low to the
high energy regime. For one single bosonic string in
$d=25$ we have that $\widetilde{\Omega}\lp E, V, 1\rp$ is given
by \cite{otros, 1997}

\begin{equation}
\widetilde{\Omega}_1(E) = \frac{V \pi^{-25/2}}{2^{24}\Gamma\lp 25/2\rp}\,
\sum_{k=0}^{\infty}\sum_{r=1}^{\infty}\,r^{-26}\,a_k^2\, E \lp
E^{\,2} - \frac{4 k r^2}{\alpha'}\rp^{23/2} \, \theta\,\lp E^{\2} -
\frac{4kr^2}{\alpha'}\rp
\end{equation}
Here, the tachyonic contribution with $k= -1$ has been suppressed. This
result amounts to summing upon the mass levels of the string for which the
masses are $m_k =\sqrt{\frac{4k}{\alpha'}}$ with $k$ a zero or natural
number. The coefficients $a_k^2$ measure the degeneration of the
corresponding mass level.

Now if $k\neq 0$, the sum over $r$ cannot be analytically performed. For
the massless contribution we have a equation of state given by
$PV=\frac{\zeta\lp 26\rp}{\zeta\lp 25\rp}\frac{N}{\beta}$. But now,
$\zeta\lp 26\rp/\zeta\lp 25\rp = 1 - 1.49\cdot 10^{-8} \sim 1$. With this
approximation, that is very natural, we are making classic the gas of massless
modes in 25 dimensions. This amounts also to taking only into account the
$r=1$ term in the second sum because the second term with $r=2$ would be
suppressed by a factor $2^{-26} \sim 1.40\cdot 10^{-8}$. We can then make
the approximation of using MB statistics to get

 \begin{equation}
\widetilde{\Omega}_1(E) = \frac{V \pi^{-25/2}}{2^{24}\Gamma\lp 25/2\rp}\,
\sum_{k=0}^{\infty}\,a_k^2\, E \lp
E^{\,2} - \frac{4 k}{\alpha'}\rp^{23/2} \, \theta\,\lp E^{\2} -
\frac{4k}{\alpha'}\rp
\end{equation}

Then, one has $\widetilde{\Omega}_1 = \Omega\lp E,1\rp$
and, in general, $\widetilde{\Omega}_n = \Omega\lp E,n\rp$. Now the average
number of objects can be computed to be

\begin{equation}
\overline{N}(E) =\frac{\sum_{n=0}^{\infty}\, n {\Omega}_n\lp
E\rp}{\sum_{n=0}^{\infty}\,{\Omega}_n\lp E\rp}
\label{nclassic}
\end{equation}
When there is a number $n=n^*$ that gives a maximum  contribution to both
sums, one obtains $\overline{N} = n^*$. So, the next step would be to
compute ${\Omega}_n$ for a very big $n$ from the single string density of
states. This can, in principle, be done by convolution, but the integral to
compute ${\Omega}_2$ is already too complicated to perform it analytically.
It seems more clever to get the behavior of ${\Omega}_1\lp E\rp$ at high
energy from the asymptotic behavior of $a_k \sim
\frac{1}{\sqrt{2}}a_0 (k-1)^{-27/4}\,
\mathrm{e}^{4 \pi \sqrt{k-1}}$. When computing ${\Omega}_2 = (1/2!)
\int_{{\scriptscriptstyle 0}}^{{\scriptscriptstyle E}}\,dt\,\Omega_1\lp
t\rp\,\Omega_1\lp E-t\rp$ we have that when $E\leq 2$ there is no massive
mode present and the convolution can be evaluated to give

\begin{equation}
\Omega_2^{\,l} (E) = \frac{1}{2 ! \Gamma\lp 50\rp}\lp \frac{24^2\Gamma\lp
25\rp}{2^{24} \pi^{25/2} \Gamma\lp 25/2\rp}\rp^2\, V^2 E^{\,49}
\end{equation}
where $l$ stands for low energy that actually means that there are only
massless modes. After going up thorough $E=2$ increasing the energy the
massive levels appear. One can define $S_2= \mathrm {ln}\,\omegat_2$ and
from here one gets $1/T_2 = \beta_2 = \partial S_2/\partial E$ which gives a
sort of fictitious temperature for the system with two strings shearing an
amount of energy $E$. Of course, the real microcanonical temperature will be
$\overline{T}_n (E)$ for big $n$ that, after substituting $\overline{N}$ as a function of
the volume an energy, will give the temperature as a function only of volume
and energy.
$\overline{T}_2(E)$ is linear with the energy for the low energy phase. The massive
modes start to curve the straight line in such a way that after passing
$\overline{T}_2=\overline{T}_H$ there appears a maximum (see Fig. \ref{t2} below). This
behavior is something normal whenever a new channel opens in any multiple
channel relativistic theory because energy is spent in mass and not in the
kinetic energy of the objects in the system. What is stringy is the fact
that no phase of positive specific heat reappears because the increasing
degeneration of the fat string prevents it. Equipartition is broken favoring
the existence of a heavy string \cite{1997}. We can use this to get an
approximate form for $\Omega_n$. Let us suppose that we have a gas that at
low temperature is a gas of massless modes. For this system is

\begin{equation}
\overline{T}_n(E) = \frac{E}{25n-1} \approx \frac{E}{25\,n}
\label{tn}
\end{equation}
where the approximation is natural because the number of convolutions is
supposed to be very big.

The Hagedorn temperature is reached at an energy $E_n$ such that
$\overline{T}_H= \frac{E_n}{25n-1}$. For energies bigger than this one, $n-1$ strings
in the gas have an energy $E_{n-1} = \overline{T}_H\lp 25n-26\rp$ and the fat string
takes $(E- E_{n-1})$ with $E\gg E_{n-1}$. So our model to approximate the
behavior of the system can be expressed in the following way \cite{npbchm}

\begin{equation}
\Omega_n\lp E\rp \approx \begin{cases}
\Omega_n^{\,l}(E)\equiv
\frac{1}{\Gamma\lp 25n\rp\,n!}\lp \frac{V\,24^{\2}\Gamma\lp 25\rp}{2^{\,24}
\pi^{25/2}\Gamma\lp 25/2\rp}\rp^{\,n}\,E^{\,25n -1}& E < E^{\,c}_n\\
\Omega_n^{\,h}(E)\equiv 
C_n\,V^n\lp E- E_{n-1}\rp^{-27/2}\,\mathrm{e}^{\,\beta_H\lp E -E_{n-1}\rp}
 & E
> E^{\,c}_n
\end{cases}
\label{approx}
\end{equation}
where $C_n$ is a function of the number of strings chosen so as to get a
continuous curve in $E^{\,c}_n$, the transition point, which belongs to the
transition region in which our approximation is not pretty good and then the
election is rather arbitrary. The important point is that this intersection
point depends on $n$.  For $n \gg 1$ we take

\begin{eqnarray*}
C_n &=&\lp \frac{\Gamma\lp 25\rp 24^{\2}}{2^{\,24}\,\pi^{25/2}\Gamma\lp
25/2\rp}\rp^n \, \frac{\lp 4\beta_H\rp^{-27/2}\,
E_{n-1}^{\,\,25n-1}}{\Gamma\lp
25n\rp\,n!\lp 5400 n\rp^{-27/4}}\\
E^{\,c}_n &=& \frac{25 n}{\beta_H}
\end{eqnarray*}

From this approximation it is easy to get Fig. \ref{t2} in which the
numerical computation of the fictitious temperature $\overline{T}_2(E)$ is
compared with $\overline{T}_2(E)$ obtained from our approximation for
$\Omega_n(E)$ in \eqref{approx} for $n=2$.

\begin{figure}[htp]
\let\picnaturalsize=N
\def\picsize{2.3in}
\def\picfilename{extfig1.eps}
\ifx\nopictures Y\else{\ifx\epsfloaded Y\else\input epsf \fi
\let\epsfloaded=Y
\centerline{\ifx\picnaturalsize N\epsfxsize \picsize\fi
\epsfbox{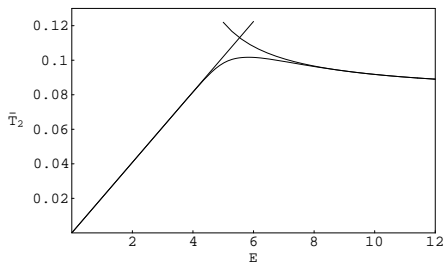}}}\fi
\caption{$\overline{T}_2(E)$ numerically computed compared with our
approximation} 
\label{t2} 
\end{figure}

\section{The average number of strings}

To compute $\overline{N}(E)$ we are going to use the approximation of the
maximum term for the sums in \eqref{nclassic}. However, now, in the sum, we
get two different functions. When energy is big, most of the terms of the
sum will correspond to the low energy regime as long as $n > n_0 =\beta_H
E/25$, but when energy is high enough $\Omega_n (E)$ will be described by
the high energy approximation for all $n < n_0$. It is then convenient to
distinguish the high energy function $\Omega_n^{\,h} (E)$ from the low
energy one,
$\Omega_n^{\,l}(E)$ as they are defined in \eqref{approx}. For the low energy
function we have, for the term with maximum contribution

\begin{equation}
n^* = \overline{N}^{\,l}(E) = D_{25}^{1/26}\,E^{25/26}\,V^{1/26}
\end{equation}
where, to simplify the equations, it has been taken $ D_{25} \equiv 
\frac{\Gamma(25) 24^{\2}}{2^{\2 4}\,\pi^{25/2}\,\Gamma\lp
25/2\rp \, 25^{25}}$ and the temperature is given by \eqref{tn}. It is then
straightforward to get the temperature as a function only of volume and
energy in a way perfectly compatible with the classical blackbody but
adapted to the polarizations of the massless states of the closed bosonic
string and the number of spatial dimensions. The Hagedorn temperature is
then reached from the low energy phase at an energy 

\begin{equation}
E_H = D_{25}\, 25^{\26}\, V\,\beta_H^{-26}        
\end{equation}
and the averaged number of objects at $E_H$ is 

\begin{equation}
N_H= 25^{\2 5} D_{25} V \beta_H^{-25} = \beta_H\,E_H/25
\label{NH}
\end{equation}

\begin{figure}[htp]
\let\picnaturalsize=N
\def\picsize{2.2in}
\def\picfilename{extfig2.eps}
\ifx\nopictures Y\else{\ifx\epsfloaded Y\else\input epsf \fi
\let\epsfloaded=Y
\centerline{\ifx\picnaturalsize N\epsfxsize \picsize\fi
\epsfbox{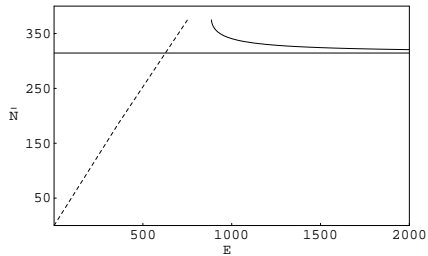}}}\fi
\caption{$\overline{N}(E)$ vs. the energy for a given volume. The dashed line
corresponds to $\overline{N}^{\,\,l}(E)$ and the continuous line 
to $\overline{N}^{\,\,h}(E)$}
\label{ene}
\end{figure}

On the other hand, the computation of $n^*$ for $\Omega_n^{\,h}(E)$ gives it
as a solution of
\br
\mathrm{ln}\,\lp V\,D_{25}\,25^{25}\rp - \mathrm{ln}\,n^* - 25
\mathrm{ln}\beta_H + \frac{27}{4 n^*} +
\frac{27}{2}\frac{25}{\beta_H\lp E - E_{n^* -1}\rp} = 0
\er
To solve this equation we use that $n^*$ will be very big and that $E \gg 25
n^*/\beta_H$ which means that we are really in the high energy regime. One
has to find the roots for a second order polynomial. One of the solutions is
not compatible with the assumption that we are in the high energy regime and
the other one gives

\begin{equation}
\overline{N}^h(E) = n^* = \frac{\beta_H}{50}\lc\,E  + E_H - \sqrt{\lp E -E_H\rp^2 -
\frac{1350}{\beta_H}\,E_H}\,\rc
\end{equation}
It is immediate to get that
$\lim_{E\rightarrow\infty}\,\overline{N}^{\,h}(E) = N_H$ (here
$E\rightarrow\infty$ physically means $E\gg E_H$). Fig. \ref{ene} shows the
average number of objects for the low and high energy regimes as a function
of $E$ for a given volume that it is fixed giving a value to $E_H$ (we also
take $\alpha' = 1$ to get the plot).
The straight line we see for the low energy part is the graphical result of
the fact that $\overline{N}^{\,l}(E)\sim E^{\,25/26} \approx E$. The part of
$C_n$ that survives in this computation is precisely what gives the
degeneration of the objects that form the gas that remains at a temperature
$\overline{T}_H$. It is easy to check that for energies lower than $E_H$ the maximum
$n^*$ really appears in the low energy region and for energies above $E_H$
it does in the high energy part. Finally it is easy to get from our
computation the form of the density of states for the high energy regime as
the result of approximating the infinite sum giving $\Lambda\lp E, V, 0\rp$
by the single contribution corresponding to
$n^*\approx N_H$

\begin{equation} 
\Omega^{\,h}(E, V)= \beta_H^{-25/2}\,\lp 5400\,N_H\lp
V\rp/16\rp^{27/4}\,\lp E-E_H\rp^{-27/2}
 \mathrm{e}^{N_H(V) + \beta_H\,E}
\end{equation}
that is compatible with what was computed longtime ago. We have made
explicit to the reader that there is a dependence on the volume for
$N_H$ as shown in \eqref{NH}. The entropy at high energy then contains terms
of the form $\frac{27}{4}\,\mathrm{ln}\,V$ and $-\frac{27}{2}\,\mathrm{ln}\,E$
that break thermodynamic extensivity because, for systems with
$\mu=0$, the extensivity of entropy amounts to the fact that $S(E, V)$,
which depends only on energy and volume after substituting $\overline{N}(E,
V)$, must be a homogeneous function of degree one of both variables. Those
terms are added to others of the form $\beta_H\,E$ and
$\beta_H^{-25}\,V$ that preserve extensivity.  This means that the limiting
system obtained at very high volume and energy, which has an infinite
calorific capacity, preserves thermodynamic extensivity. The term
$-\frac{27}{2}\,\mathrm{ln}\,E$ is actually responsible for $1/C_V$ to be 
different from zero and, in fact, negative. This is the way we
thermodynamically recover what we have physically implemented, as the result
of the breaking of equipartition, in the statistical mechanics construction
of $\Omega_N$. In it, from the very beginning, we have distinguished the
single fat string from the sea and then the procedure is inherently non
extensive.

\section{Comments}

Following \cite{vb}, the negative specific heat for the gas of closed
strings that appears in the microcanonical description when no winding modes
are present has been promoted to a selection principle ruling out those
systems. We do not know what the basics to prefer the situation in which the
specific heat is positive are. Moreover, when it is used to the
extent of automatically ruling out every model for which, in a
microcanonical description, a negative specific heat appears. Perhaps what
has happened is that the fact that no negative specific heat can appear in
any canonical (fixed temperature) description (based in Boltzmann-Gibbs
statistics) has been unjustifiably transfered from canonical thermodynamics
to the microcanonical one. Anyhow, what is well established \cite{gross} is
that the microcanonical entropy $S$ and the microcanonical inverse
temperature $\overline{\beta}$ are single valued, multiple differentiable,
and smooth. Furthermore, as $\beta > 0$ and
$\partial^{\2}S/\partial E^{\2}\neq
\pm\infty$, the microcanonical $C_V$ can be positive or negative but never
zero. In conclusion, a priori  there is nothing wrong with having negative
microcanonical $C_V(E)$. Furthermore, from a thermodynamic point of view,
the occurrence of a first-order phase transition in the canonical ensemble
defined by the presence of a non differentiable point of the free energy may
appear as a non concavity of the entropy as a function of energy and,
consequently, as a situation of non equivalence of ensembles \cite{can}. Several
empirical systems in atomic and condensed matter physics are also known for
which negative specific heat has been measured (cf. for example
\cite{schmidt}). In any case, as for strings, the negative specific heat for
these systems is related to a loss of thermodynamic extensivity. However,
for the laboratory examples, non extensivity results from the presence of
long-range interactions when the system is of any size or short-range
interactions in the case it is a small one.

Non equivalence of ensembles is a topic also related to the appearance of
fluctuations \cite{carlitz}. This is the case for the big fluctuations that
emerge in the canonical internal energy near Hagedorn and that imply the
need to distinguish canonical and microcanonical descriptions. Fluctuations
in the number of particles are also related to the possibility of
approaching both sums in \eqref{nbar1} for the single contribution given by
the same $N=N^*$.

A possible alternative to try to understand the Hagedorn phase is to think
that we really have an almost out-of-equilibrium phenomenon for which a
sort of generalized statistics like the one in \cite{tsallisreyni} has to be
used. It is immediate to see that such attempt implies the use of a factor
different from the Boltzmann factor $\mathrm{e}^{\,-\beta\,H}$ and that would
certainly change (may be suppress) the Hagedorn behavior in the canonical
descriptions. The problem is that, as lucidly exposed in \cite{nauenberg},
there are very serious objections against this kind of non extensive
generalized statistics.

\section*{Acknowledgments}

M.A.C. acknowledges partial support by a Spanish MCYT-FPI fellowship. M. S.
is partially supported by a Spanish MEC-FPU fellowship. We all are partially 
supported in our work by the MCYT project BFM2003-00313/FISI.

\end{document}